\documentclass[prb, twocolumn, showpacs, superscriptaddress]{revtex4-1}
\usepackage{amsmath}
\usepackage{amssymb}
\usepackage{hyperref}
\usepackage[normalem]{ulem}
\usepackage{graphicx}
\usepackage{color}

\begin{document}

\title{Exceptional Points in a Non-Hermitian Topological Pump}

\author{Wenchao Hu}

\thanks{These two authors contributed equally to this work.}

\affiliation{Centre for Disruptive Photonic Technologies, Nanyang Technological University, Singapore 637371, Singapore}
\affiliation{School of Electrical and Electronic Engineering, Nanyang Technological University, 50 Nanyang Avenue, 639798, Singapore}

\author{Hailong Wang}

\thanks{These two authors contributed equally to this work.}

\affiliation{Division of Physics and Applied Physics, School of Physical and Mathematical Sciences, Nanyang Technological University, Singapore 637371, Singapore}

\author{Perry Ping Shum}

\affiliation{Centre for Disruptive Photonic Technologies, Nanyang Technological University, Singapore 637371, Singapore}
\affiliation{School of Electrical and Electronic Engineering, Nanyang Technological University, 50 Nanyang Avenue, 639798, Singapore}

\author{Y.~D.~Chong}
\email{yidong@ntu.edu.sg}

\affiliation{Centre for Disruptive Photonic Technologies, Nanyang Technological University, Singapore 637371, Singapore}
\affiliation{Division of Physics and Applied Physics, School of Physical and Mathematical Sciences, Nanyang Technological University, Singapore 637371, Singapore}

\begin{abstract}
We investigate the effects of non-Hermiticity on topological pumping,
and uncover a connection between a topological edge invariant based on
topological pumping and the winding numbers of exceptional points.  In
Hermitian lattices, it is known that the topologically nontrivial
regime of the topological pump only arises in the infinite-system
limit.  In finite non-Hermitian lattices, however, topologically
nontrivial behavior can also appear.  We show that this can be
understood in terms of the effects of encircling a pair of exceptional
points during a pumping cycle.  This phenomenon is observed
experimentally, in a non-Hermitian microwave network containing
variable gain amplifiers.
\end{abstract}


\maketitle

\section{Introduction}

The existence of topologically distinct phases of matter was one of
the most profound discoveries of theoretical physics in recent
decades.\cite{KaneRMP} The idea of classifying bandstructures using
topological invariants, such as the Chern number,\cite{Thouless1982}
arose originally in the study of the quantum Hall effect,\cite{MStone}
and subsequently led to the discovery of two- and three-dimensional
topologically insulating materials.  It has also inspired numerous
proposals and experiments for realizing topologically non-trivial
bands using light,\cite{Soljacic2014} sound,\cite{Baile2015,Alu2015}
and other types of waves.\cite{Derek2012} According to the topological
bulk-edge correspondence principle,\cite{Bernevig} topologically
nontrivial bandstructures imply the existence of
topologically-protected edge states, whose unique transport properties
may have applications in many fields.

Theories of bandstructure topology typically assume that the
underlying lattice is Hermitian.  Yet in settings such as topological
photonics, non-Hermitian effects---in the form of gain and/or
loss---are easily introduced, and may be both substantial and
unavoidable in practical implementations.\cite{Baile2016} Broadly
speaking, non-Hermiticity poses two problems for standard theories.
Firstly, non-Hermitian bands can exhibit exceptional points
(EPs),\cite{Kato, Heiss2012} in which case bands cease to be
continuously single-valued throughout $k$-space, which is conceptually
troublesome for band invariants such as Chern numbers.  Secondly,
standard formulations of the bulk-edge correspondence principle rely
on the existence of a real spectral gap in the bulk.  For instance, in
Hatsugai's well-known derivation of the correspondence between quantum
Hall edge states and Chern numbers, it is crucial to assume that a
lattice in a strip geometry has a \textit{real} point spectrum which
converges, in the large-system limit, to an integer number of real
bands.\cite{Hatsugai}

A number of recent works\cite{Levitov, Esaki, Bardyn, CTChan2015,
  TonyLee, Leykam} have started to explore how band topological ideas
might be extended to non-Hermitian systems.  Notably, certain special
one- and two-dimensional non-Hermitian lattices have been found to
exhibit topological invariants with meaningful bulk-edge
correspondences.\cite{TonyLee,Leykam} Some of these novel invariants
are based on the integral winding numbers associated with EPs of the
complex band spectrum (i.e., branch point orders), which constitute a
natural class of discrete features tied to non-Hermiticity.  So far,
however, it has been unclear whether there is a connection between
invariants based on EP winding numbers and the standard topological
invariants previously developed for Hermitian systems.

This paper demonstrates, theoretically and experimentally, a
relationship between a previously-known Hermitian topological
invariant and EP winding numbers.  The topological invariant comes
from a \textit{topological pump},\cite{Laughlin, Brouwer0, Brouwer,
  Fulga, Longwen} in the form of the winding numbers of scattering
matrix eigenvalues during a parametric pumping cycle.\cite{Yidong2015}
In a Hermitian lattice, the bulk-edge correspondence principle
associates a zero (nonzero) winding number with a topologically
trivial (nontrivial) bulk bandgap.  Strictly speaking, however,
topologically nontrivial behavior emerges only in the
$N\rightarrow\infty$ limit, where $N$ is the sample width (i.e., the
limit where opposite edges of the sample are decoupled).  We show that
when the topological pump is continued into the non-Hermitian case,
e.g.~by applying both gain and loss, topologically nontrivial behavior
can arise under the generalized condition
\begin{equation}
  |\gamma| \gtrsim e^{-N},
  \label{gamma_constraint}
\end{equation}
where $\gamma$ parameterizes the non-Hermiticity (gain/loss) in each
unit cell.  In the Hermitian case ($|\gamma| \rightarrow 0$),
satisfying (\ref{gamma_constraint}) requires taking $N
\rightarrow\infty$.  For a non-Hermitian lattice ($\gamma \ne 0$),
nonzero windings can occur for finite $N$, and we show that these
emerge out of the winding numbers of a pair of EPs of the non-unitary
scattering matrix.

The above ideas are realized and confirmed in an experiment on a
classical electromagnetic network in a topological pumping
configuration,\cite{Laughlin, Brouwer0, Brouwer} operating in the
microwave frequency range (900 MHz).  Previously, we have shown that
such a network, constructed from radio frequency (RF) cables,
couplers, and phase shifters, can be used to implement a topological
pump.\cite{Yidong2015} The windings of the scattering matrix
eigenvalues, during a pumping cycle, were found to match the
underlying bandstructure of the network, which could be topologically
trivial or nontrivial.  In that experiment, the intrinsic losses were
fixed.\cite{Yidong2015} Here, we incorporate \textit{tunable}
amplifiers that can be used to control the level of non-Hermiticity.
This allows us to probe for the theoretically-predicted EPs, and study
their effects on the topological pump.

\section{Exceptional points and topological pumping}
\label{sec:Theory}

\begin{figure}
  \centering       
  \includegraphics[width=\columnwidth]{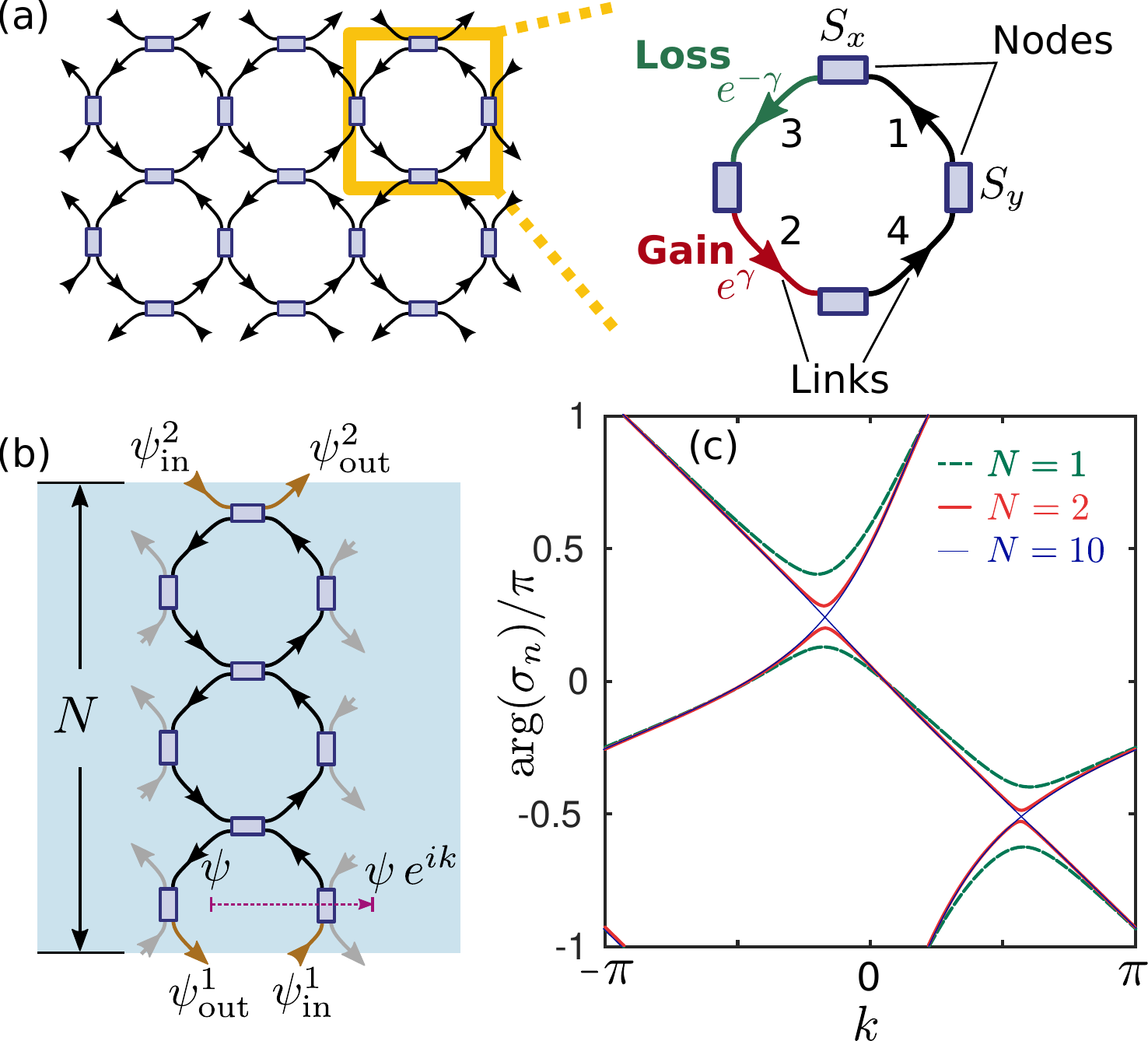}
  \caption{(a) Schematic of a periodic network of directed links and nodes, with 
  	tunable gain and loss in two of the links in each unit cell.  (b) Schematic of 
  	the setup for the Laughlin-Brouwer topological pump.  The network is 
  	truncated to $N$ unit cells in the $y$ direction, and twisted boundary 
  	conditions are applied in the $x$ direction with tunable twist angle $k$ 
  	(corresponding to an infinite strip with fixed wavenumber).  (c) Plot of 
  	$\mathrm{arg} (\sigma_n)$ versus $k$, where $\{\sigma_n\}$ are the 
  	eigenvalues of the edge scattering matrix.  The model parameters are 
  	$\gamma = 0$ (Hermitian network), $\theta_x=\theta_y=3\pi/8$,
  	$\Phi^3_x=\Phi^3_y=-7\pi/10$, and $\phi=2\pi/5$; these parameters are 
  	defined in Appendix~\ref{2Dnetwork}.}
  \label{fig:pump_schematic}
\end{figure}

We begin with a theoretical analysis of a 2D square-lattice network of
directed links joined by nodes, shown in
Fig.~\ref{fig:pump_schematic}(a).  Steady-state waves propagating in
the links are described by complex scalar amplitudes; at each node,
the input and output amplitudes are related by a $2\times2$ coupling
matrix.  As shown in previous works,\cite{Pasek2014} if all links have
line delay $\phi$, wave propagation in an infinite periodic network is
described by the Floquet equation
 \begin{equation}
   U(\mathbf k)|\psi(\mathbf k)\rangle=e^{-i\phi} |\psi(\mathbf k)\rangle.
  \label{eigval_prob}
 \end{equation}
For a Bloch state with real crystal wave-vector $\mathbf k$, 
$|\psi(\mathbf k)\rangle$ denotes a vector of wave amplitudes exiting the 
nodes of one unit cell in the network, and $U(\mathbf k)$ describes the 
``scattering'' of the wave by the unit cell.  Eq.~(\ref{eigval_prob}) thus describes 
a Floquet bandstructure, with $\phi(\mathbf k)$ as a quasienergy.

The network is ``Hermitian'' if there is no gain or loss, so that all
propagation and scattering processes are unitary.  Then $U(\mathbf k)$
is unitary, and the Floquet Hamiltonian $H_F(\mathbf k) =
i\log[U(\mathbf k)]$ is Hermitian.  Moreover, the quasienergies
$\phi(\mathbf k)$ are real, and the bandstructure can be topologically
trivial or non-trivial, depending on system parameters like nodal
coupling strengths.  (The topologically non-trivial phase is an
``anomalous Floquet insulator'', with interesting properties that have
been studied in previous works.)\cite{Demler0, Demler, Derek2014,
  HL2015, Lindner, Pasek2014, HL2016}

A topological edge invariant can be formulated for the network by
truncating it in the $y$ direction to form a strip $N$ cells wide.
Dirichlet boundary conditions are imposed on the upper and lower
edges, so that no waves enter or leave via these edges.  Translational
invariance in $x$ gives a conserved wave-number $k$, equivalent to
taking one unit cell along $x$ and imposing twisted boundary
conditions.  One can then calculate the bandstructure $\phi(k)$, and
count the net number of edge states on each edge and in each bandgap.
This is a topological invariant, independent of the choice of
truncation direction for the strip.\cite{Hatsugai}

Another way to formulate a topological edge invariant is the
Laughlin-Brouwer topological pump.\cite{Laughlin, Brouwer0, Brouwer}
As shown in Fig.~\ref{fig:pump_schematic}(b), instead of imposing
Dirichlet boundary conditions on the edges, we attach scattering
leads.  The wave amplitudes incident on the two edges,
$|\psi_{\mathrm{in}}\rangle$, can be related to the output amplitudes
$|\psi_{\mathrm{out}}\rangle$ by
 \begin{equation}
   |\psi_{\mathrm{out}}\rangle
   =S_{\mathrm{edge}}|\psi_{\mathrm{in}}\rangle\;.
  \label{scattering}
 \end{equation}
The derivation of $S_{\mathrm{edge}}$ is described in 
Appendix~\ref{2Dnetwork}.  If the network is Hermitian, $S_{\mathrm{edge}}$ 
is unitary and its eigenvalues $\{\sigma_n\}$ lie on the complex unit circle.  To 
perform topological pumping, we set $\phi$ in a bandgap and advance $k$ by 
$2\pi$, and then count the resulting winding number of the $\sigma_n$'s along 
the unit circle.\cite{Brouwer0,Brouwer}

\begin{figure}
  \centering       
  \includegraphics[width=1.0\linewidth]{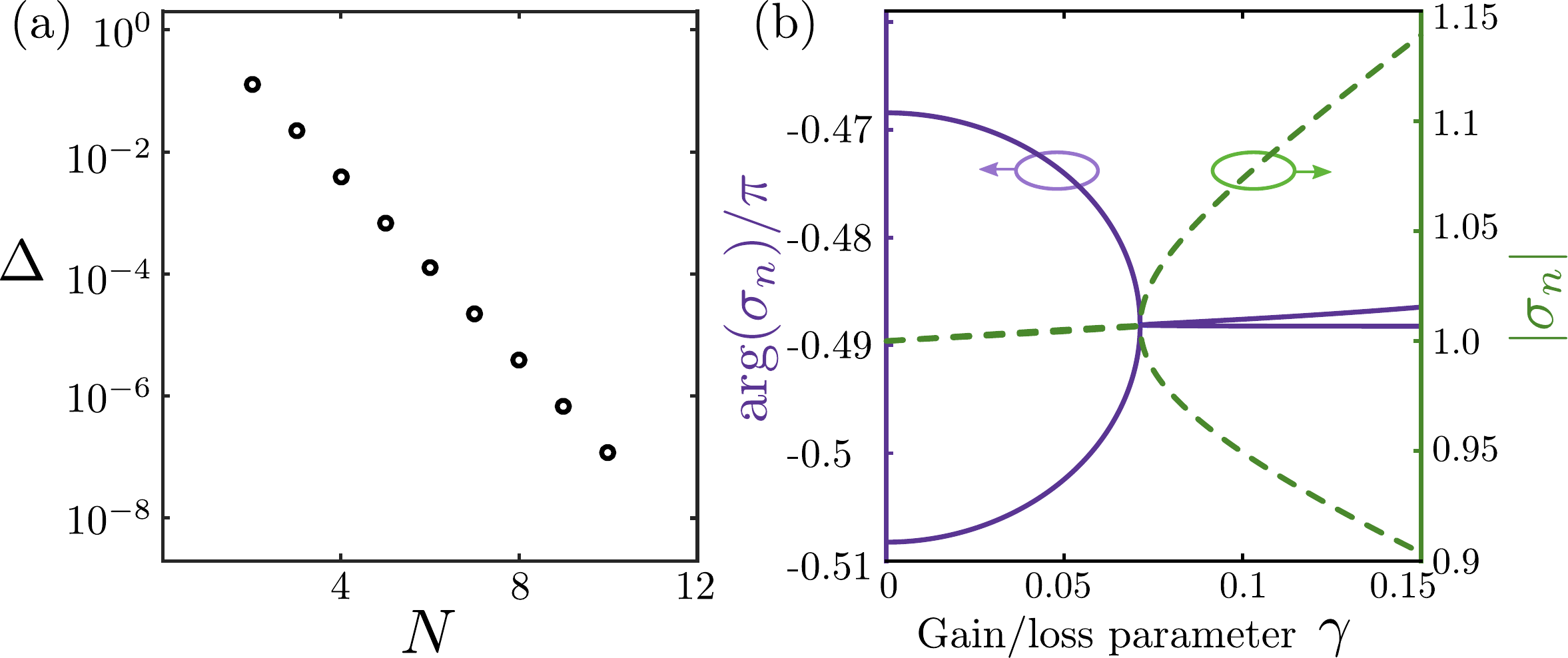}
  \caption{ (a) The gaps between arguments of scattering matrix
    eigenvalues,
    $\Delta\equiv\text{min}\left(|\mathrm{arg}(\sigma_1)-\mathrm{arg}
    (\sigma_2)|\right)$, in the Hermitian limit $\gamma=0$.  Circles
    represent gaps for different $N$.  (b) Scattering matrix
    eigenvalues of the non-Hermitian 2D network, at $k_{ep}\simeq
    0.572\pi$ and $N=2$.  The left and right axes show the arguments
    and amplitudes of the eigenvalues, respectively.  The other model
    parameters are same as in Fig.~\ref{fig:pump_schematic} (c):
    $\theta_x=\theta_y=3\pi/8$, $\Phi^3_x=\Phi^3_y=-7\pi/10$, and
    $\phi=2\pi/5$.}
  \label{fig:gap_and_EP}
\end{figure}

Fig.~\ref{fig:pump_schematic}(c) plots $\mathrm{arg}(\sigma_n)$ versus
$k$ for various strip widths $N$ for the Hermitian network.  All the
other system parameters are chosen so that the underlying bulk
bandstructure is topologically nontrivial and $\phi\in\mathbb{R}$ lies
in a bandgap.  For small $N$, the individual eigenvalues have
\textit{zero} winding around the origin during one cycle of the
pumping parameter $k$.  For larger $N$, the points of nearest
separation between the eigenvalues appear to shrink.  To study this in
greater detail, Fig.~\ref{fig:gap_and_EP}(a) plots
$\Delta\equiv\mathrm{min}\big|\mathrm{arg}[\sigma_1(k)]-\mathrm{arg}
[\sigma_2(k)]\big|$, where the minimum is calculated over $k \in
[-\pi,\pi]$, for various finite $N$.  This quantifies the minimum
separation between the two eigenvalue trajectories, and must vanish if
the eigenvalues cross somewhere in $k \in [-\pi,\pi]$ (which is
required for nonzero windings).  These numerical results show that
$\Delta \sim \exp(-N)$, reaching zero only for $N \rightarrow \infty$.
Physically, this reflects the fact that topological protection is
spoiled by the coupling of edge states on opposite edges of the
sample; since edge state wavefunctions decay exponentially into the
bulk, the coupling strength decreases exponentially with $N$.

What is the effect of the above topological pumping process on a
non-Hermitian network?  To study this problem, we focus on
non-Hermitian networks with a specific distribution of gain and loss,
shown in Fig.~\ref{fig:pump_schematic}(a): in each unit cell, link 2
has gain factor $\exp(\gamma)$, link 3 has loss factor
$\exp(-\gamma)$, and the other links remain unitary.  Thus, $\gamma$
simultaneously tunes gain and loss in links 2 and 3.  This arrangement
of gain and loss is chosen so that exceptional points (EPs) of the
system are easily accessible, and affect the behavior of the
topological pump.  (Note that it is not $\mathcal{P}\mathcal{T}$
(parity-time) symmetric;\cite{Bender1998, Bender2002, Kostas2008,
  Yidong2011, Rechtsman2015} in Appendix~\ref{PTsymmetry}, we show
that in an alternative $\mathcal{P}\mathcal{T}$ symmetric version of
the network, the gain/loss distribution does not affect the
topological pump.)

We can fix $\phi$ and $\gamma$, and carry out the ``pumping''
procedure as before: the parameter $k$ is advanced by $2\pi$, and we
examine how the trajectories of $\{\sigma_n\}$---the eigenvalues of
the non-unitary $S_{\mathrm{edge}}$ matrices---wind in the complex
plane.  Note that the variation of $k$ is a parametric evolution, not
a time evolution, so the breakdown of adiabaticity in non-Hermitian
systems\cite{Moiseyev} is not an issue.  We now observe an interesting
feature of the non-Hermitian pump: nonzero windings can occur for
\textit{finite} $N$, due to the existence of exceptional points (EPs)
of $S_{\mathrm{edge}}$.

An EP is a point in a 2D parameter space where a matrix becomes
defective and its eigenvectors become linearly dependent.\cite{Kato,
  Heiss2012} Due to the spectral theorem, EPs only appear in
non-Hermitian systems.  In Fig.~\ref{fig:gap_and_EP}(b), we plot
$\mathrm{arg}(\sigma_n)$ and $|\sigma_n|$ against the gain/loss
parameter $\gamma$, for $N = 2$ and $k = 0.572\pi$.  The two
eigenvalues exhibit bifurcative behavior characteristic of an EP,
coalescing at $\gamma = 0.071$.  In this case, $S_{\mathrm{edge}}$
possesses a pair of EPs in the 2D parameter space formed by $k$ and
$\gamma$; one EP is located at $(k = 0.572\pi, \gamma = 0.071)$, as
seen in Fig.~\ref{fig:gap_and_EP}(b), while the other EP is located at
$(k = -0.1731\pi, \gamma = 0.1132)$.

\begin{figure}
  \centering       
  \includegraphics[width=\columnwidth]{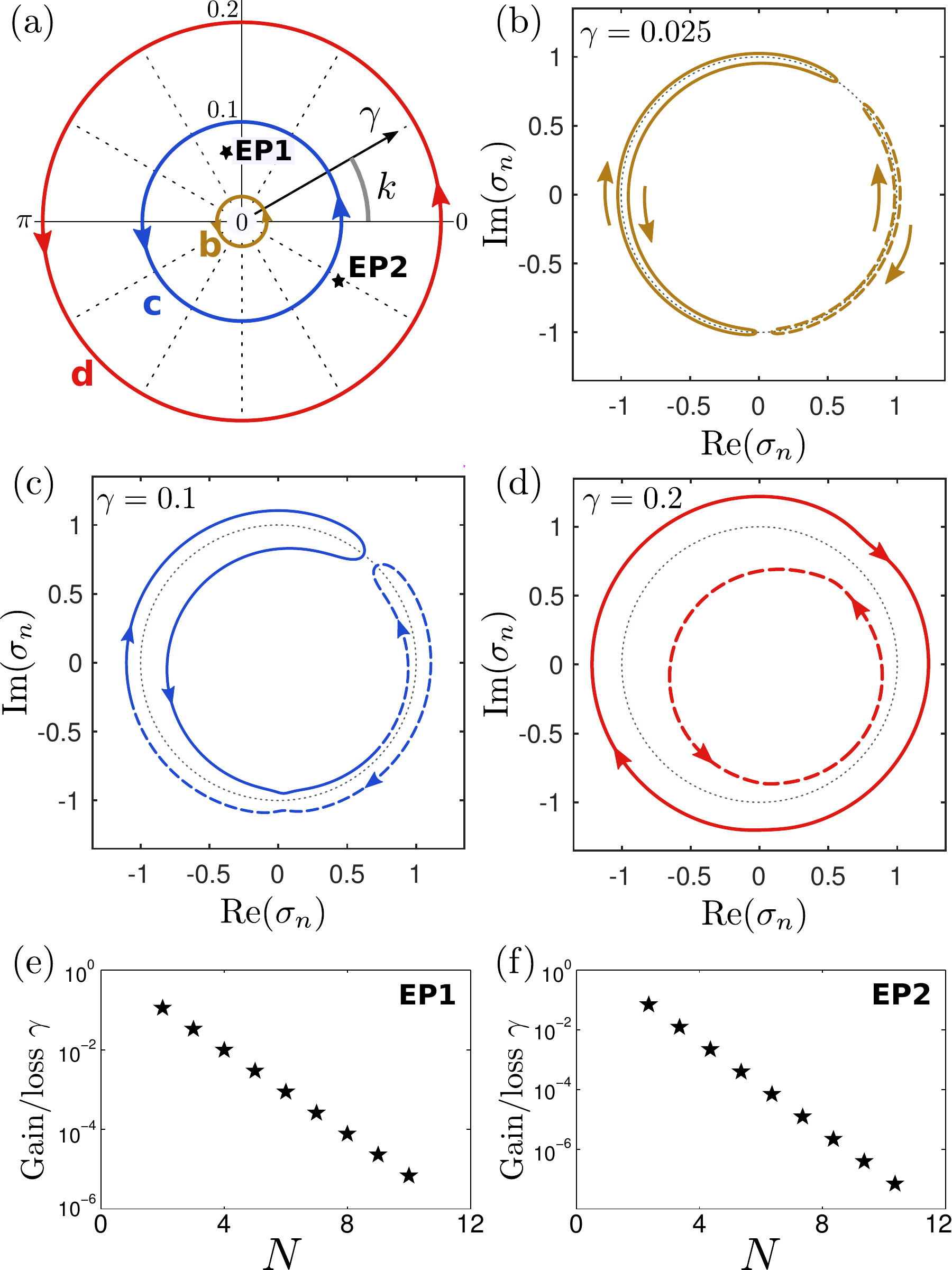}
  \caption{Relationship between topological pumping and exceptional
    points, calculated for a network model of width $N=2$.  The
    underlying bandstructure is topologically nontrivial, with $\phi$
    in a bandgap ($\theta_x=\theta_y=3\pi/8$,
    $\Phi^3_x=\Phi^3_y=-0.7\pi$, and $\phi=2\pi/5$).  (a) Parametric
    loops in the 2D parameter space formed by the gain/loss parameter
    $\gamma$ (radial coordinate) and pumping parameter $k$ (azimuthal
    coordinate).  Exceptional points (EPs) of $S_{\mathrm{edge}}$ are
    indicated by stars.  (b, c, d) Complex plane trajectories of
    $\{\sigma_n\}$, the eigenvalues of $S_{\mathrm{edge}}$,
    corresponding to the parametric loops shown in (a).  The two
    distinct eigenvalue trajectories are plotted as solid and dashed
    curves, and the unit circle is plotted as dots for comparison.
    The trajectories in (b) have zero winding around the origin, like
    the Hermitian finite-$N$ limit; the trajectories in (c) join each
    other under one cycle, because the parametric loop encloses one
    exceptional point; the trajectories in (d) have nonzero windings,
    similar to the Hermitian large-$N$ limit.  (e)--(f) Plots of the
    gain/loss parameter $\gamma$ at the two EPs [as labelled in (a)],
    for different $N$.  }
  \label{fig:topology1}
\end{figure}

Fig.~\ref{fig:topology1} illustrates how the EPs give rise to the
topologically nontrivial regime of the topological pump.
Fig.~\ref{fig:topology1}(a) shows the 2D parameter space, with the
gain/loss parameter $\gamma$ serving as the radial coordinate and the
cyclic pumping parameter $k$ serving as the azimuthal coordinate.  The
two EPs of $S_{\mathrm{edge}}$ are labeled EP1 and EP2 (these EP
positions depend on $N$, which is set here to $N=2$).  In
Fig.~\ref{fig:topology1}(b)--(d), we plot the trajectories of
$\{\sigma_n\}$ in the complex plane, as the system proceeds along the
different parametric loops indicated in Fig.~\ref{fig:topology1}(a).
For a parametric loop at small $\gamma$, not enclosing any EP, the
eigenvalue trajectories do not wind around the origin, as shown in
Fig.~\ref{fig:topology1}(b).  This behavior extends down to the
previously-discussed Hermitian limit ($\gamma = 0$).  For a parametric
loop at large $\gamma$, enclosing both EPs, the eigenvalue
trajectories have nonzero windings, as shown in
Fig.~\ref{fig:topology1}(d), and is similar to the large-$N$ limit of
the Hermitian topological pump.  Between these two regimes, there are
two points in the parameter space where the eigenvalue trajectories
cross, which are EPs of $S_{\mathrm{edge}}$.
Fig.~\ref{fig:topology1}(c) shows the intermediate regime where only
one EP is enclosed by the parametric loop.  In this case, one pumping
cycle induces a continuous exchange of the two eigenvalues, along with
their eigenvectors.\cite{Heiss2012}

\begin{figure}
  \centering       
  \includegraphics[width=1.0\linewidth]{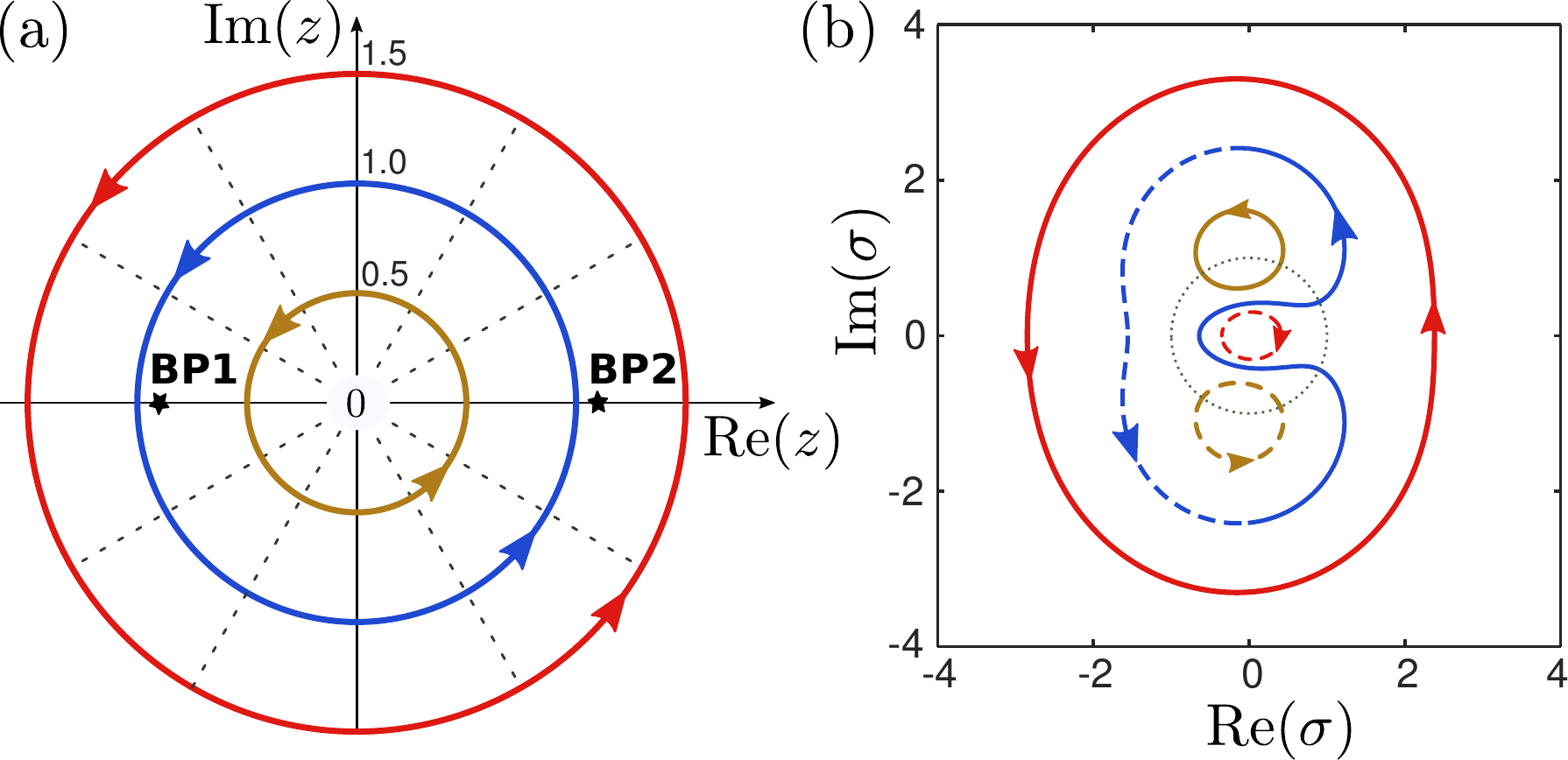}
  \caption{Behavior of the multi-valued function $\sigma(z)$ given by
    Eq.~(\ref{complex_model}), illustrating how branch points can
    produce winding and non-winding trajectories.  Here, we take
    $\alpha = 0.1$.  (a) Plot of the parameter space formed by the
    complex variable $z$, with the branch points of $\sigma(z)$
    indicated by stars, along with three parametric loops
    corresponding to (i) $|z| = 0.5$, (ii) $|z| = 1$, and (iii)
    $|z|=1.5$.  (b) The corresponding complex plane trajectories of
    the two branches of $\sigma(z)$. }
  \label{fig:EP_illu}
\end{figure}

We can use a simple model to illustrate how such a mathematical relationship 
between EPs and eigenvalue windings might arise.  Consider the multi-valued 
function
\begin{equation}
  \sigma(z) = z - \alpha + \sqrt{(z-\alpha+1)(z-\alpha-1)}.
  \label{complex_model}
\end{equation}
The two branches of $\sigma(z)$, arising from the square root, could
represent solutions to a secular equation for the eigenvalues of a
$2\times2$ matrix, parameterized analytically by a variable $z$; the
branch points, $z = \alpha \pm 1$, are EPs of that matrix.
Fig.~\ref{fig:EP_illu}(a) shows three different loops in the parameter
space, centered at $z = 0$ and enclosing zero, one, and two branch
points, similar to Fig.~\ref{fig:topology1}(a).  In
Fig.~\ref{fig:EP_illu}(b), we plot the trajectories of $\sigma(z)$
corresponding to those parametric loops, and observe winding behaviors
very similar to Fig.~\ref{fig:topology1} (b)--(d).  In particular, for
$|z| \rightarrow 0$, branches of Eq.~(\ref{complex_model}) reduce to
$\sigma_\pm(z) \approx - \alpha\pm \sqrt{\alpha^2-1}$, exhibiting zero
winding during one cycle of $\mathrm{arg}(z)$; whereas for
$|z|\rightarrow \infty$, the branches reduce to $\sigma_\pm(z) =
(2z)^{\pm1}$, which wind in opposite directions around the origin.

These findings imply that the topologically nontrivial regime of the
Hermitian topological pump\cite{Laughlin,Brouwer0,Brouwer} emerges
from the general behavior of the non-Hermitian topological pump, via
an appropriate order of limits.  For given finite $N$, nonzero
windings require sufficiently large $\gamma$, as expressed in
Eq.~(\ref{gamma_constraint}).  Fig.~\ref{fig:topology1}(e)--(f) shows
the values of $\gamma$ at the EPs for different $N$.  With increasing
$N$, the required level of non-Hermiticity decreases exponentially,
reaching zero for $N\rightarrow\infty$.


\section{Experiment}
\label{Experiment}

\begin{figure}
  \centering       
  \includegraphics[width=0.45\textwidth]{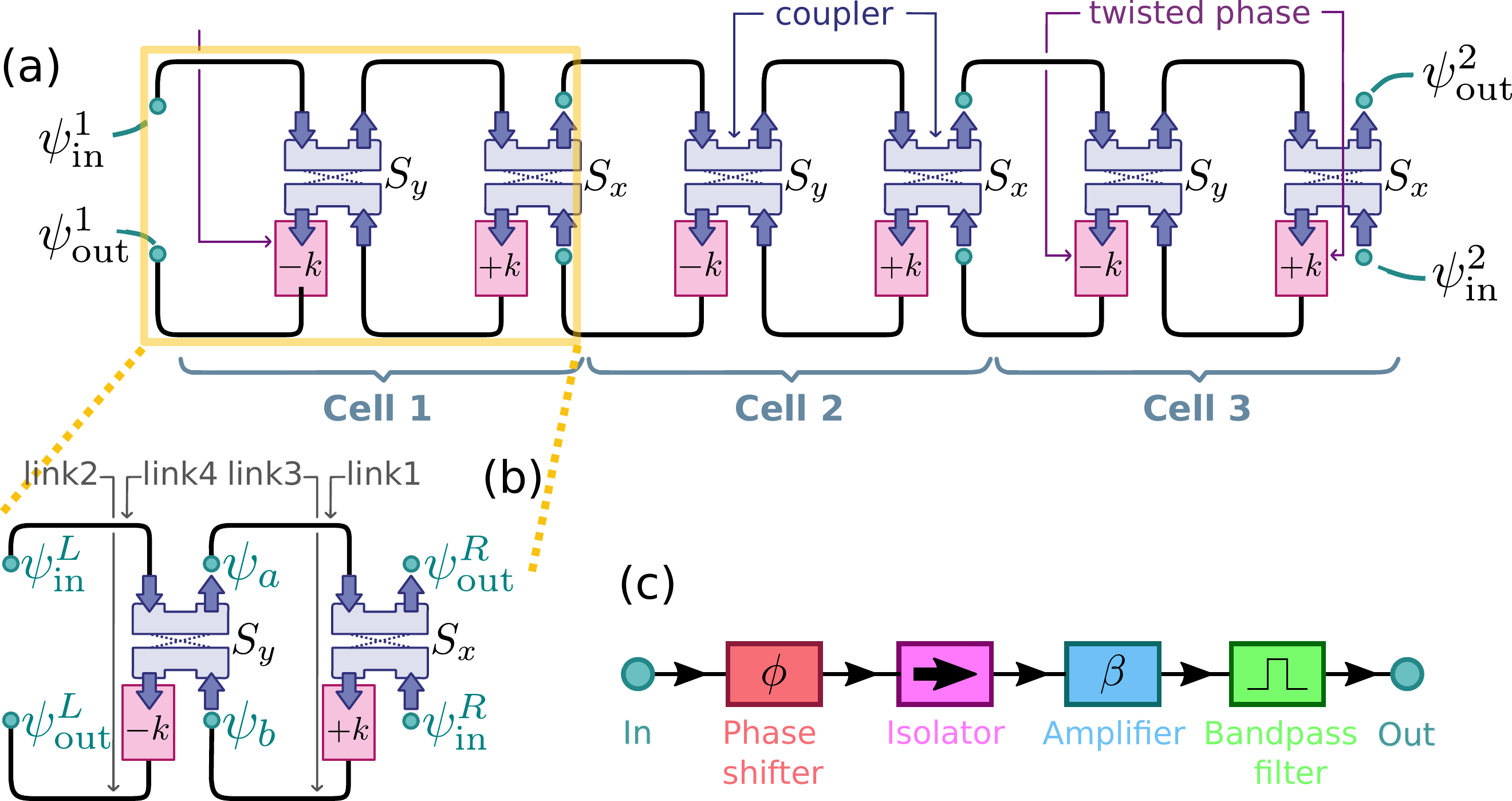}
  \caption{ (a) Experimental setup.  Each of the identical units,
    labeled ``Cell 1", ``Cell 2" and ``Cell 3," corresponds to one
    cell in the topological pump geometry.  The twisted boundary
    condition is applied by tuning phase shifter (pink boxes) in lower
    links. The couplers (blue rods) are depicted in the
    strong-coupling configuration.  The weak-coupling configuration is
    achieved by swapping each coupler's outputs. The overall input and
    output amplitudes are $\psi_{\mathrm{in}}^{1,2}$ and
    $\psi_{\mathrm{out}}^{1,2}$. Their scattering parameters are
    measured with a network analyzer. (b) Each cell is composed of
    four links ($j=1,2,3,4$) and two couplers ($S_x, S_y$).  (c) Every
    link in our system are exactly the same and contain one phase
    shifter, one isolator, one digital controlled variable gain
    amplifier, one bandpass filter and five low-loss handflex
    interconnect coaxial RF cables. }
  \label{fig:network}
\end{figure}

In order to realize the non-Hermitian topological pump, we implemented the 
model described in Section~\ref{sec:Theory} using a classical electromagnetic 
network operating at microwave (900\,MHz) frequencies.  The basic setup is 
shown in Fig.~\ref{fig:network}, and is conceptually similar to the experiment 
previously reported in Ref.~\onlinecite{Yidong2015}.  The network is designed 
according to the topological pumping configuration shown in 
Fig.~\ref{fig:pump_schematic}(b).  It corresponds to a ``column'' of the periodic 
network of width $N$, composed of identical unit cells each containing four 
links and two nodes.

Each directional link consists of five low-loss hand-flex coaxial RF cables 
(086-10SM+/086-15SM+, Mini-Circuits), a bandpass filter (CBC-893+, Mini-
Circuits), an isolator (S0091IAD, Nova Microwave), a phase shifter (SPHSA-152+, 
Mini-Circuits), and a digitally-controlled variable-gain amplifier (DVGA1-242+, 
Mini-Circuits).  The link's transmission coefficient, $t = \beta \exp(i\varphi)$, can 
be independently tuned in both phase and amplitude.  The phase $\varphi
\in[0, 2\pi)$ is controlled by the phase shifter with $\pm1^\circ$
  precision, and is used to set the network model quasienergy $\phi$
  and the pumping parameter $k$.  The gain/loss factor $\beta$ is
  tunable in the range $[-8\,\text{dB}, \,24\,\text{dB}]$ with $\pm
  0.25\,\text{dB}$ precision.  The $-8\,\text{dB}$ lower bound
  corresponds to turning off the amplifier, and comes from the
  intrinsic losses of the components (which are substantially lower
  than in Ref.~\onlinecite{Yidong2015}, due in part to the lower
  operating frequency of 900\,MHz rather than 5\,GHz.)

Each node consists of an RF coupler (BDCN-7-25+, Mini-Circuits) with
$\approx 1\!:\!7$ coupling ratio.  At 900\,MHz, its measured $S$
parameters are
\begin{equation}
  S_{\text{coupler}}=
  \begin{bmatrix}
    0.914e^{-i0.622\pi} & 0.348e^{-i0.127\pi} \\
    0.348e^{-i0.127\pi} & 0.914e^{-i0.622\pi}
  \end{bmatrix}.
  \label{eq:s_coupler}
\end{equation}
By swapping the order of the output ports, we can realize both
topologically trivial and nontrivial phases of the network model's
underlying bandstructure.\cite{Yidong2015} In terms of the coupling
parameter $\theta$ defined in Appendix~\ref{2Dnetwork}, this means we
can set $\theta \approx 0.12\pi$ (weak-coupling/trivial) or $\theta
\approx 0.38\pi$ (strong-coupling/nontrivial), where $\theta = 0.5\pi$
is the topological transition point.

\begin{figure}
  \centering       
  \includegraphics[width=0.45\textwidth]{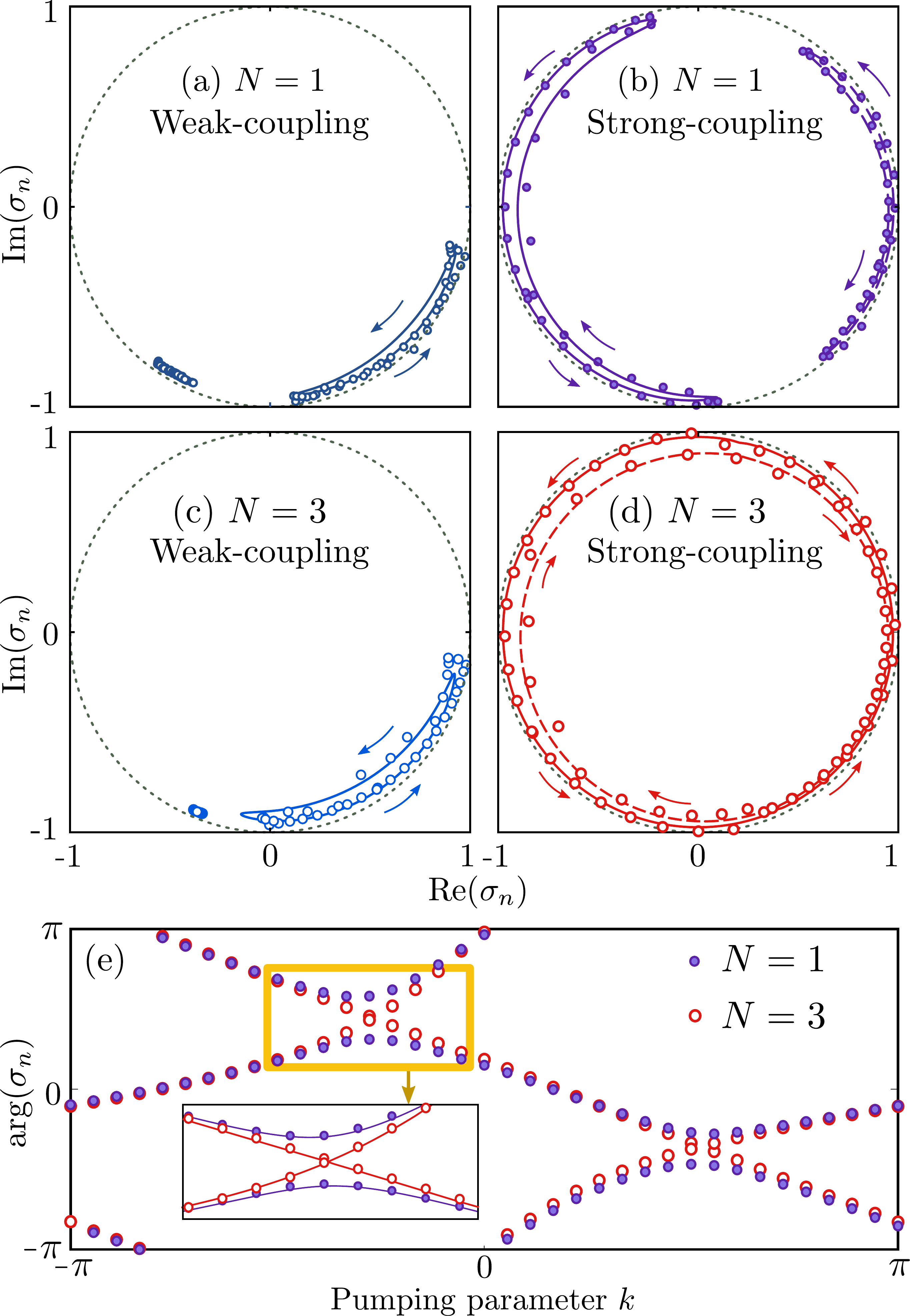}
  \caption{ (a)--(d) Experimentally measured scattering matrix
    eigenvalues over one pumping cycle, with approximately zero
    gain/loss.  Results are shown for (a,b) $N=1$ and (c,d) $N=3$.
    The weak-coupling regime corresponds to a topologically trivial
    phase of the underlying network bandstructure, while the
    strong-coupling regime corresponds to a topologically nontrivial
    phase.  Directly measured data is plotted with dots, and
    calculations using measured network-component S-parameters are
    shown as solid and dashed curves.  The unit circle is indicated by
    dotted curves. (e) Arguments of the measured scattering matrix
    eigenvalues, $\mathrm{arg}(\sigma_n)$, versus pumping parameter
    $k$.  The network is in a topologically nontrivial phase.  Inset:
    behavior near a crossing point; solid curves show theoretical
    results calculated using the network components' measured
    $S$-parameters, which predict an avoided crossing for $N=1$ and a
    crossing for $N=3$. }
  \label{fig:eigenvalue_ideal}
\end{figure}

To measure the edge scattering matrix, $S_{\mathrm{edge}}$, we attach
the inputs and outputs at the ends of the ``column'' to a vector
network analyzer (Anritsu 37396C).  Fig.~\ref{fig:eigenvalue_ideal}
shows the results when $\beta \approx 1$ (i.e., with the amplifiers
are tuned so that there is no net gain or loss in each link).  In this
case, $S_{\mathrm{edge}}$ is approximately unitary, and as expected
the eigenvalues lie very close to the complex unit circle.  Under one
cycle of $k$, we observe no winding in the weak-coupling case.  In the
strong-coupling case, nonzero windings appear only when the system
size is sufficiently large, as shown in
Fig.~\ref{fig:eigenvalue_ideal}(d).  This is consistent with the
results reported in Ref.~\onlinecite{Yidong2015}, and with the
theoretical discussion of Section~\ref{sec:Theory}.

\begin{figure} 
  \centering       
  \includegraphics[width=0.49\textwidth]{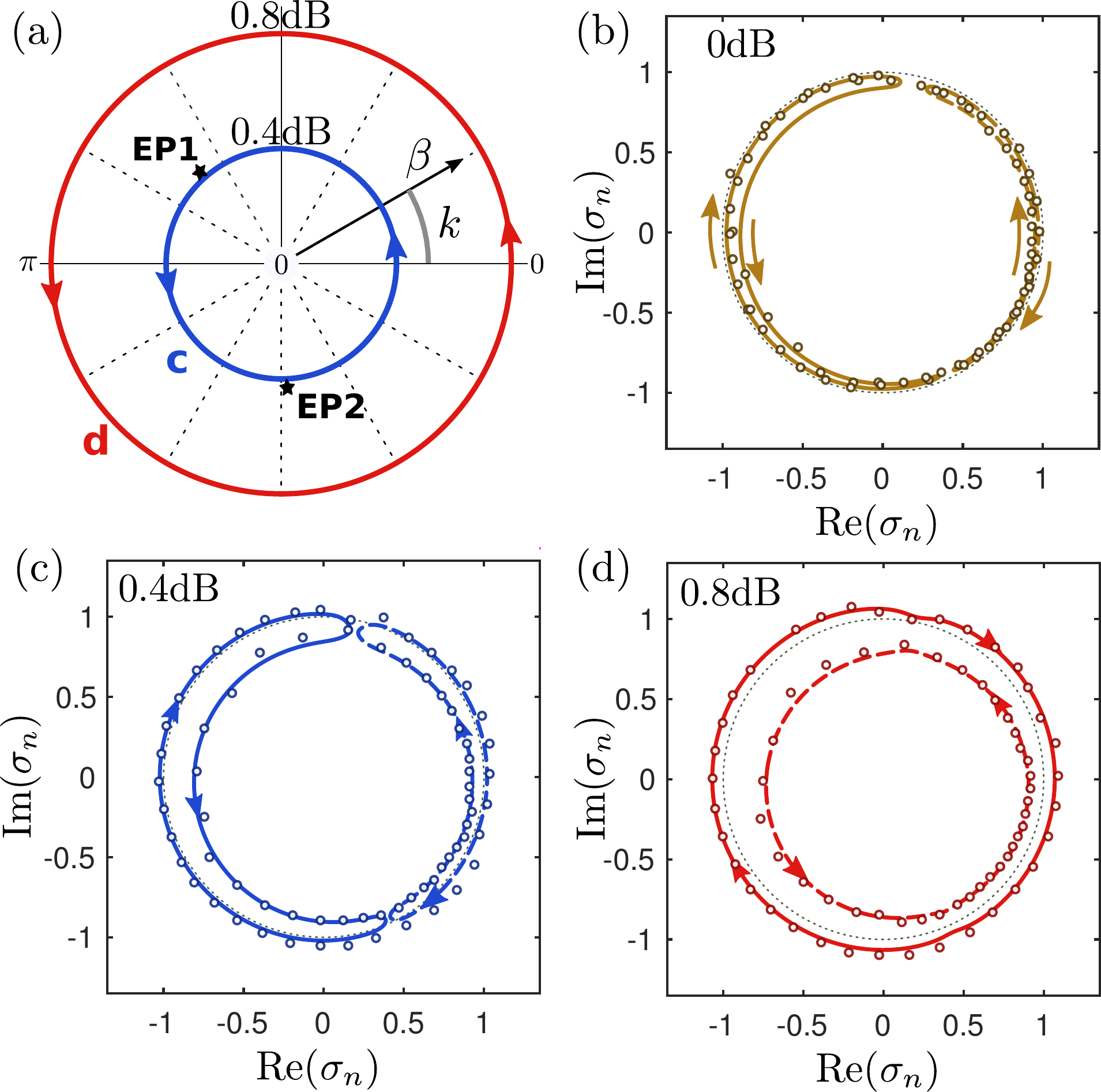}
  \caption{ (a) Parametric loops in the 2D parameter space formed by
    the logarithmic gain/loss factor $\beta$ (radial coordinate) and
    pumping parameter $k$ (azimuthal coordinate).  The gain/loss
    distribution is as depicted in Fig.~\ref{fig:network}(a).
    Exceptional point positions (stars) are calculated from the
    measured S-parameters of the network components.  The system size
    is $N=2$.  (b)--(d) Eigenvalues of $S_{\mathrm{edge}}$ in the
    complex plane.  Directly measured data is plotted with dots, and
    calculations using measured network-component S-parameters are
    shown as solid and dashed curves.  The unit circle is indicated by
    dotted curves. }
  \label{fig:eigenvalue_gain}
\end{figure}  

Next, we implement the explicitly non-Hermitian topological pumping
configuration discussed in Section~\ref{sec:Theory}.  To do this, we
selectively apply amplification to some of the links, according to the
gain/loss distribution shown in Fig.~\ref{fig:pump_schematic}(a).  The
results, for $N = 2$, are shown in Fig.~\ref{fig:eigenvalue_gain}.
The parameter space is depicted in Fig.~\ref{fig:eigenvalue_gain}(a);
our calculations, based on the measured S-parameters of the individual
network components, indicate that there are two EPs very close to
$\beta=0.4$dB.  We take three different parametric loops, with results
shown in Fig.~\ref{fig:eigenvalue_gain}(b)--(d).  In
Fig.~\ref{fig:eigenvalue_gain}(c), the parametric loop passes very
close to both EPs, and we observe the two eigenvalue trajectories
nearly meeting at two bifurcation points, similar to
Fig.~\ref{fig:gap_and_EP}(b).  In Fig.~\ref{fig:eigenvalue_gain}(d),
when both EPs are enclosed by the parametric loop, the eigenvalue
trajectories acquire nonzero windings.

\section{Summary}
\label{summary}

We have performed a theoretical and experimental study of a
non-Hermitian topological pump.  We have based our study on a network
model, which has previously been shown to be a convenient and
experimentally feasible way to realize a topological pump using
classical microwaves~\cite{Yidong2015}.  When gain and loss is added
to the network, we find that topologically nontrivial behavior
(nonzero windings) requires the pumping process to encircle two
exceptional points in the parameter space of the non-Hermitian system.
This criterion is a generalization of the $N \rightarrow \infty$ limit
which is necessary for topological protection to emerge in Hermitian
topological pumps.\cite{Laughlin, Brouwer0, Brouwer} These theoretical
ideas were demonstrated experimentally, using a microwave network
containing variable-gain amplifiers.  In future work, we seek to
generalize this finding to a wider class of non-Hermitian lattices,
not necessarily described by network models.

\section{Acknowledgements}

We thank Longwen Zhou, Chunli Huang and Daniel Leykam for helpful
comments.  This research was supported by the Singapore National
Research Foundation under grant No.~NRFF2012-02, by the Singapore MOE
Academic Research Fund Tier 2 Grant No. MOE2015-T2-2-008, and by the
Singapore MOE Academic Research Fund Tier 3 grant MOE2011-T3-1-005.


\appendix

\section{Scattering matrix of the 2D network}
\label{2Dnetwork}

This appendix describes the derivation of the scattering matrix for
the network model discussed in Section \ref{sec:Theory}.  The network
is truncated in the $y$ direction, to form a strip $N$ cells wide.
The periodicity in $x$ yields a wave-number $k$.  We can equivalently
regard this as a supercell of $N$ unit cells, with twisted boundary
conditions along the $x$ boundaries with twist angle $k$.  The
supercell can be further divided into $N$ identical subunits, each
composed of four links ($j=1,2,3,4$) and two couplers ($S_x, S_y$), as
shown in Fig.~\ref{fig:network}(b).

At any given node in the supercell, the incoming and outgoing wave
amplitudes are related by a $2\times2$ unitary coupling relation
 \begin{equation}
   \begin{pmatrix}
      \psi_{\text{out}}^{R} \\ 
      e^{-ik-\gamma_3-i\phi}\psi_b \\
   \end{pmatrix}
   =S_x
   \begin{pmatrix}
      e^{\gamma_1+i\phi}\psi_a \\ 
      \psi_{\text{in}}^{R} \\
   \end{pmatrix}\;.
  \label{eqn:scattering1}
 \end{equation}
The coupling matrices at the nodes have the simple form
 \begin{equation}
   S_x=
   \begin{pmatrix}
      \sin\theta_x & i\cos\theta_x \\ 
      i\cos\theta_x & \sin\theta_x \\
   \end{pmatrix},
\end{equation}
corresponding to couplers with $180^\circ$ rotational symmetry, where
$\theta_x$ is the coupling parameter in the $x$ direction.  The
discussion could also be generalized to arbitrary $2\times2$ unitary
coupling matrices of the form
\begin{equation*}
S_{\mu}
    =e^{i\Phi^3_\mu}
  \begin{pmatrix}
    \sin\theta_\mu e^{-i(\Phi^1_\mu + \Phi^2_\mu)}
    &i\cos\theta_\mu e^{i(\Phi^1_\mu - \Phi^2_\mu)} \\
    i\cos\theta_\mu e^{-i(\Phi^1_\mu - \Phi^2_\mu)}
    &\sin\theta_\mu e^{i(\Phi^1_\mu + \Phi^2_\mu)}\\
  \end{pmatrix},
\end{equation*}
where $\{\Phi_1^{\mu},\Phi_2^{\mu},\Phi_3^{\mu}\}$ are additional
Euler angles.  Apart from some phase shifts in the wave amplitudes,
these additional Euler angles may cause systematic shifts in $k$ and
$\phi$; however, such shifts do not alter the topological properties
of the network's bandstructure.

The coupling relation at the other node is given by
 \begin{equation}
   \begin{pmatrix}
      e^{ik-\gamma_2-i\phi}\psi_{\text{out}}^{L} \\ \psi_a \\ 
   \end{pmatrix}
   =S_y
   \begin{pmatrix}
      e^{\gamma_4+i\phi}\psi_{\text{in}}^{L} \\ \psi_b \\
   \end{pmatrix}\;.
  \label{eqn:scattering2}
 \end{equation}
We can use Eqs.~(\ref{eqn:scattering1}) and (\ref{eqn:scattering2}) to
obtain an analytic relation of the form
 \begin{equation}
   \begin{pmatrix}
      \psi_{\text{out}}^{R} \\ \psi_{\text{in}}^{R} \\ 
   \end{pmatrix}
   =M
   \begin{pmatrix}
      \psi_{\text{in}}^{L} \\ \psi_{\text{out}}^{L} \\ 
   \end{pmatrix}\;,
  \label{eqn:transfer1a}
 \end{equation}
where $M$ is the transfer matrix for one subunit.  For a supercell of $N$
subunits, we have
 \begin{equation}
   \begin{pmatrix}
      \psi_{\text{out}}^{2} \\ \psi_{\text{in}}^{2} \\ 
   \end{pmatrix}
   =\tilde M
   \begin{pmatrix}
      \psi_{\text{in}}^{1} \\ \psi_{\text{out}}^{1} \\ 
   \end{pmatrix}\;,
 \end{equation}
where $\tilde M\equiv M^N$ is the total transfer matrix for the
supercell.  We can show that
 \begin{equation*}
   \tilde M=\frac{\lambda_2^N-\lambda_1^N}{\lambda_2-\lambda_1}M 
   -\frac{\lambda_2^{N-1}-\lambda_1^{N-1}}
   {\lambda_2-\lambda_1}\det{(M)}\;,
 \end{equation*}
where $\lambda_{1,2}$ are eigenvalues of the transfer matrix $M$:
\begin{equation*}
\lambda_{1,2}=\frac{\mathrm{Tr}(M)}{2}\pm\sqrt{\left[\frac{\mathrm{Tr}(M)}{2}\right]^2-\det{(M)}}\;.
\end{equation*}

The outgoing and incoming wave amplitudes can then be related via the
scattering matrix relation
 \begin{equation}
   \begin{pmatrix}
      \psi_{\text{out}}^{1} \\ \psi_{\text{out}}^{2} \\ 
   \end{pmatrix}
   =S_{\mathrm{edge}}
   \begin{pmatrix}
      \psi_{\text{in}}^{1} \\ \psi_{\text{in}}^{2} \\ 
   \end{pmatrix}\;,
  \label{eqn:scattering_basis}
 \end{equation}
where 
 \begin{equation}
   S_{\mathrm{edge}}=\frac{1}{\tilde M_{22}}
   \begin{pmatrix}
      -\tilde M_{21} & 1 \\ \det{(\tilde M)} & \tilde M_{12} \\
   \end{pmatrix}\;.
  \label{eqn:scattering_matrix}
 \end{equation}

\section{$\mathcal{P}\mathcal{T}$-symmetric 2D network}
\label{PTsymmetry}

As mentioned in Section \ref{sec:Theory}, the non-Hermitian network
that we have studied has gain and loss in certain links, but this
gain/loss distribution is not $\mathcal{P}\mathcal{T}$
(parity/time-reveral) symmetric.\cite{Bender1998, Bender2002,
  Kostas2008, Yidong2011, Rechtsman2015} Alternatively, it is possible
for us to impose a $\mathcal{P}\mathcal{T}$ symmetric gain/loss
distribution on the network.  Referring to
Fig.~\ref{fig:pump_schematic}(a), this can be accomplished by adding
balanced gain and loss to links 1 and 3, and to links 2 and 4: i.e.,
$\gamma_1=-\gamma_3$ and $\gamma_4=-\gamma_2$.

In that case, the topological pump's $S_{\mathrm{edge}}$ matrix will
satisfy
\begin{equation}
  \mathcal{P}\mathcal{T}
  S_{\mathrm{edge}}(-k,-\phi,-\gamma)
  \mathcal{P}\mathcal{T} =
  S^{-1}_{\mathrm{edge}}(k,\phi,\gamma),
  \label{eqn:PTsymmetry}
\end{equation}
where $\mathcal{P}$ is the first Pauli matrix ($\sigma_x$) and
$\mathcal{T}$ is the complex conjugation operator.
Eq.~(\ref{eqn:PTsymmetry}) is closely analogous to the symmetry
relation obeyed by scattering matrices derived from
$\mathcal{P}\mathcal{T}$ symmetric wave
equations.\cite{Yidong2011,Rechtsman2015}

For this $\mathcal{P}\mathcal{T}$ symmetric network, it can be shown
that $M\propto \exp(\gamma_1+\gamma_4)$ and $\tilde M\propto
\exp[N(\gamma_1+\gamma_4)]$.  The scattering matrix eigenvalues
$\{\sigma_n\}$ are thus independent of $\gamma_1$ and $\gamma_4$.  The
system behaves like a Hermitian topological pump, regardless of the
level of non-Hermiticity.

\end{document}